%% file: journal_paper.tex
\documentclass[12pt,3p,12pt,review]{elsarticle}

\usepackage[utf8]{inputenc}
\usepackage[active]{srcltx}
\usepackage{bm}
\usepackage{hhline}
\usepackage{amsmath,amsthm,amssymb}
\usepackage[linesnumbered,ruled,vlined]{algorithm2e}

\makeatletter
\usepackage{amssymb}
\usepackage{amsmath}
\usepackage{caption}
\usepackage{subcaption}
\usepackage{booktabs}
\usepackage{url}
\usepackage{array}
\usepackage[flushleft]{threeparttable}
\usepackage{longtable}
\usepackage{setspace}
\usepackage{multirow}
\usepackage[usenames, dvipsnames]{color}
\usepackage{xcolor}
\usepackage{rotating}
\usepackage{tikz}
\usepackage{pgfplots}
\usepackage{pgfkeys}
\usetikzlibrary{spy,calc}

\newif\ifblackandwhitecycle
\gdef\patternnumber{0}

\pgfkeys{/tikz/.cd,
    zoombox paths/.style={
        draw=orange,
        very thick
    },
    black and white/.is choice,
    black and white/.default=static,
    black and white/static/.style={ 
        draw=white,   
        zoombox paths/.append style={
            draw=white,
            postaction={
                draw=black,
                loosely dashed
            }
        }
    },
    black and white/static/.code={
        \gdef\patternnumber{1}
    },
    black and white/cycle/.code={
        \blackandwhitecycletrue
        \gdef\patternnumber{1}
    },
    black and white pattern/.is choice,
    black and white pattern/0/.style={},
    black and white pattern/1/.style={    
            draw=white,
            postaction={
                draw=black,
                dash pattern=on 2pt off 2pt
            }
    },
    black and white pattern/2/.style={    
            draw=white,
            postaction={
                draw=black,
                dash pattern=on 4pt off 4pt
            }
    },
    black and white pattern/3/.style={    
            draw=white,
            postaction={
                draw=black,
                dash pattern=on 4pt off 4pt on 1pt off 4pt
            }
    },
    black and white pattern/4/.style={    
            draw=white,
            postaction={
                draw=black,
                dash pattern=on 4pt off 2pt on 2 pt off 2pt on 2 pt off 2pt
            }
    },
    zoomboxarray inner gap/.initial=5pt,
    zoomboxarray columns/.initial=2,
    zoomboxarray rows/.initial=2,
    subfigurename/.initial={},
    figurename/.initial={zoombox},
    zoomboxarray/.style={
        execute at begin picture={
            \begin{scope}[
                spy using outlines={%
                    zoombox paths,
                    width=\imagewidth / \pgfkeysvalueof{/tikz/zoomboxarray columns} - (\pgfkeysvalueof{/tikz/zoomboxarray columns} - 1) / \pgfkeysvalueof{/tikz/zoomboxarray columns} * \pgfkeysvalueof{/tikz/zoomboxarray inner gap} -\pgflinewidth,
                    height=\imageheight / \pgfkeysvalueof{/tikz/zoomboxarray rows} - (\pgfkeysvalueof{/tikz/zoomboxarray rows} - 1) / \pgfkeysvalueof{/tikz/zoomboxarray rows} * \pgfkeysvalueof{/tikz/zoomboxarray inner gap}-\pgflinewidth,
                    magnification=3,
                    every spy on node/.style={
                        zoombox paths
                    },
                    every spy in node/.style={
                        zoombox paths
                    }
                }
            ]
        },
        execute at end picture={
            \end{scope}
            \node at (image.north) [anchor=north,inner sep=0pt] {\subcaptionbox{\label{\pgfkeysvalueof{/tikz/figurename}-image}}{\phantomimage}};
            \node at (zoomboxes container.north) [anchor=north,inner sep=0pt] {\subcaptionbox{\label{\pgfkeysvalueof{/tikz/figurename}-zoom}}{\phantomimage}};
     \gdef\patternnumber{0}
        },
        spymargin/.initial=0.5em,
        zoomboxes xshift/.initial=1,
        zoomboxes right/.code=\pgfkeys{/tikz/zoomboxes xshift=1},
        zoomboxes left/.code=\pgfkeys{/tikz/zoomboxes xshift=-1},
        zoomboxes yshift/.initial=0,
        zoomboxes above/.code={
            \pgfkeys{/tikz/zoomboxes yshift=1},
            \pgfkeys{/tikz/zoomboxes xshift=0}
        },
        zoomboxes below/.code={
            \pgfkeys{/tikz/zoomboxes yshift=-1},
            \pgfkeys{/tikz/zoomboxes xshift=0}
        },
        caption margin/.initial=4ex,
    },
    adjust caption spacing/.code={},
    image container/.style={
        inner sep=0pt,
        at=(image.north),
        anchor=north,
        adjust caption spacing
    },
    zoomboxes container/.style={
        inner sep=0pt,
        at=(image.north),
        anchor=north,
        name=zoomboxes container,
        xshift=\pgfkeysvalueof{/tikz/zoomboxes xshift}*(\imagewidth+\pgfkeysvalueof{/tikz/spymargin}),
        yshift=\pgfkeysvalueof{/tikz/zoomboxes yshift}*(\imageheight+\pgfkeysvalueof{/tikz/spymargin}+\pgfkeysvalueof{/tikz/caption margin}),
        adjust caption spacing
    },
    calculate dimensions/.code={
        \pgfpointdiff{\pgfpointanchor{image}{south west} }{\pgfpointanchor{image}{north east} }
        \pgfgetlastxy{\imagewidth}{\imageheight}
        \global\let\imagewidth=\imagewidth
        \global\let\imageheight=\imageheight
        \gdef\columncount{1}
        \gdef\rowcount{1}
        
    },
    image node/.style={
        inner sep=0pt,
        name=image,
        anchor=south west,
        append after command={
            [calculate dimensions]
            node [image container,subfigurename=\pgfkeysvalueof{/tikz/figurename}-image] {\phantomimage}
            node [zoomboxes container,subfigurename=\pgfkeysvalueof{/tikz/figurename}-zoom] {\phantomimage}
        }
    },
    color code/.style={
        zoombox paths/.append style={draw=#1}
    },
    connect zoomboxes/.style={
    spy connection path={\draw[draw=none,zoombox paths] (tikzspyonnode) -- (tikzspyinnode);}
    },
    help grid code/.code={
        \begin{scope}[
                x={(image.south east)},
                y={(image.north west)},
                font=\footnotesize,
                help lines,
                overlay
            ]
            \foreach \x in {0,1,...,9} { 
                \draw(\x/10,0) -- (\x/10,1);
                \node [anchor=north] at (\x/10,0) {0.\x};
            }
            \foreach \y in {0,1,...,9} {
                \draw(0,\y/10) -- (1,\y/10);                        \node [anchor=east] at (0,\y/10) {0.\y};
            }
        \end{scope}    
    },
    help grid/.style={
        append after command={
            [help grid code]
        }
    },
}

\newcommand\phantomimage{%
    \phantom{%
        \rule{\imagewidth}{\imageheight}%
    }%
}

\newcommand\zoombox[2][]{
    \begin{scope}[zoombox paths]
        \pgfmathsetmacro\xpos{
            (\columncount-1)*(\imagewidth / \pgfkeysvalueof{/tikz/zoomboxarray columns} + \pgfkeysvalueof{/tikz/zoomboxarray inner gap} / \pgfkeysvalueof{/tikz/zoomboxarray columns} ) + \pgflinewidth
        }
        \pgfmathsetmacro\ypos{
            (\rowcount-1)*( \imageheight / \pgfkeysvalueof{/tikz/zoomboxarray rows} + \pgfkeysvalueof{/tikz/zoomboxarray inner gap} / \pgfkeysvalueof{/tikz/zoomboxarray rows} ) + 0.5*\pgflinewidth
        }
        \edef\dospy{\noexpand\spy [
            #1,
            zoombox paths/.append style={
                black and white pattern=\patternnumber
            },
            every spy on node/.append style={#1},
            x=\imagewidth,
            y=\imageheight
        ] on (#2) in node [anchor=north west] at ($(zoomboxes container.north west)+(\xpos pt,-\ypos pt)$);}
        \dospy
        \pgfmathtruncatemacro\pgfmathresult{ifthenelse(\columncount==\pgfkeysvalueof{/tikz/zoomboxarray columns},\rowcount+1,\rowcount)}
        \global\let\rowcount=\pgfmathresult
        \pgfmathtruncatemacro\pgfmathresult{ifthenelse(\columncount==\pgfkeysvalueof{/tikz/zoomboxarray columns},1,\columncount+1)}
        \global\let\columncount=\pgfmathresult
        \ifblackandwhitecycle
            \pgfmathtruncatemacro{\newpatternnumber}{\patternnumber+1}
            \global\edef\patternnumber{\newpatternnumber}
        \fi
    \end{scope}
}

\newtheorem{theorem}{Theorem}
\newtheorem{definition}{Definition}

\makeatother

\newcommand{\dottedcolumn}[3]{%
  \settowidth{\dimen0}{$#1$}
  \settowidth{\dimen2}{$#2$}
  \ifdim\dimen2>\dimen0 \dimen0=\dimen2 \fi
  \begin{pmatrix}\,
    \vcenter{
      \kern.6ex
      \vbox to \dimexpr#1\normalbaselineskip-1.2ex{
        \hbox{$#2$}
    \kern3pt
    \xleaders\vbox{\hbox to \dimen0{\hss.\hss}\vskip4pt}\vfill
    \kern1pt
    \hbox{$#3$}
  }\kern.6ex}\,
  \end{pmatrix}
}

\SetCommentSty{mycommfont}

\begin{document}
\begin{frontmatter} 

\title{Local Verlet buffer approach for broad-phase interaction detection in Discrete Element Method}

\address{
Faculty of Science, Technology and Communication, University of Luxembourg, Campus Belval, 2, avenue de l'Université, 4365 Esch-sur-Alzette, Luxembourg
}

\author[]{Abdoul Wahid Mainassara Checkaraou}
\author[]{Xavier Besseron \corref{cor1}}
\ead{xavier.besseron@uni.lu} 
\author[]{Alban Rousset}
\author[]{Fenglei Qi}
\author[]{Bernhard Peters}

\cortext[cor1]{Corresponding author}

\begin{abstract}
The Extended Discrete Element Method (XDEM) is a novel and innovative numerical simulation technique that extends the dynamics of granular materials or particles as described through the classical discrete element method (DEM) by additional properties such as the thermodynamic state, stress/strain for each particle. Such DEM simulations used by industries to set up their experimental processes are complexes and heavy in computation time.

Those simulations perform at each time step a collision detection to generate a list of interacting particles that is one of the most expensive computation parts of a DEM simulation. The Verlet buffer method, which was first introduced in Molecular Dynamic (MD) (and is also used in DEM) allows to keep the interaction list for many time step by extending each particle neighbourhood by a certain extension range, and thus broadening the interaction list. The method relies mainly on the stability of the DEM, which ensures that no particles move erratically or unpredictably from one time step to the next: this is called temporal coherency. In the classical and current approach, all the particles have their neighbourhood extended by the same value which leads to suboptimal performances in simulations where different flow regimes coexist. Additionally, and unlike in MD (which remains very different from DEM on several aspects), there is no comprehensive study analysing the different parameters that affect the performance of the Verlet buffer method in DEM.

In this work, we apply a dynamic neighbour list update method that depends on the particles individual displacement, and an extension range specific to each particle and based on their local flow regime for the generation of the neighbour list. The update of the interaction list is analysed throughout the simulation based on the particles displacement allowing a flexible update according to the flow regime conditions.
We evaluate the influence of the Verlet extension range on the performance of the execution time through different test cases and we empirically analyse and define the extension range value giving the minimum of the global simulation time.

\end{abstract}


\end{frontmatter} 


\section{Introduction}
\label{sec:introduction}
\input{introduction.tex}

\section{Related work}
\label{sec:related_work}
\input{related_work.tex}

\section{Background}
\label{sec:background_xdem}
\input{background_xdem.tex}

\section{Local Verlet buffer approach }
\label{subsec:verlet_buffer}
\input{verlet_buffer.tex}

\section{Performance Evaluation}
\label{sec:param_study}
\input{performance_eval.tex}

\section{Conclusion}
\label{sec:conclusion}
\input{conclusion.tex}

\section*{Acknowledgement}

This research is in the framework of the “PowderReg” project, funded by the European program Interreg VA GR within the priority axis 4 "Strengthen the competitiveness and the attractiveness of the Grande Region/Gro\ss region". The experiments presented in this paper were carried out using the HPC facilities of the University of Luxembourg~\cite{VBCG_HPCS14}.

\clearpage
\section*{References}

\bibliographystyle{elsarticle-num}
\bibliography{Bibliography}

\end{document}

%% file: introduction.tex
Discrete Element Method (DEM), originally proposed by Cundall~\cite{Cundall79}, is a popular simulation approach for studying and diagnosing bulk powder/granular dynamic systems, which are ubiquitous in pharmaceutical industry, food processing, chemical engineering, mining industry, and energy systems~\cite{ketterhagen2009process, ransing2000powder}. Considering the large scale of applied systems, one of the key efforts in the DEM development is to enhance the simulation capability of  DEM software, such as, by adopting advanced parallelism schemes~\cite{maknickas2006parallel}, utilizing graphics processing units (GPU)~\cite{gan2016gpu} and developing coarse grain models~\cite{weinhart2016influence}. However, one unavoidable functionality that DEM codes need to optimize is collision detection, which, including neighbour search, represents a major computational part in DEM simulations~\cite{rousset2018comparing,pall2013flexible}. 
Collision detection is often split into two phases: a broad-phase, which formulates a potential collision list for each particle (neighbor list), and a narrow-phase accounting for accurately resolving the collision instance of each pair of particles in the neighbor list. Different algorithms for constructing neighbor list in the broad-phase are available, including brute force approach~\cite{kockara2007collision} of $O(n^2)$ time complexity, binning approach~\cite{tracy2009efficient} and linked-cell method~\cite{welling2011efficiency} of $O(n)$ complexity. However, the broad-phase computational efficiency is not solely determined by the time complexity of the adopted algorithm~\cite{rousset2018comparing}, which are also affected, for instance, by the  ratio of cell size to particle size in the commonly used linked-cell approach or by the frequency of updating the neighbour list in the broad-phase. The latter is usually related to the Verlet buffer approach that is firstly proposed in MD simulations by Loup Verlet~\cite{verlet1967computer} for reducing the unnecessary cost of rebuilding neighbor list at every simulation time step. The mechanism for skipping neighbor list rebuild is achieved by providing an extra margin (often called "skin") on top of the particles pairwise cut-off interaction distance. The neighbor list built with Verlet buffer is called Verlet list. With this mechanism, the Verlet list remains unchanged until a particle displacement exceeds a certain threshold distance. \par 

 For MD simulations,  the systems are often homogeneous and a uniform (global) buffer is satisfactory to achieve a good speed-up. However, for majority of powder and granular dynamic systems, the variation of particle flow properties such as particle velocity and solid fraction in the systems is significant.  It becomes less efficient to adopt a uniform skin margin for particles at regions of different flow conditions. Intuitively, in such systems, providing a larger skin margin for particles moving faster leads to a more reasonable neighbor list updating frequency, globally.  
Many parameter studies on the skin margin determination have been reported in MD simulation research~\cite{verlet1967computer,chialvo1990use,chialvo1991performance,mattson1999near,sutmann2006optimization}.  For MD, in Lennard Jones systems, often $skin = 0.3 \sigma$ is chosen, where $\sigma$ is the diameter of a Lennard Jones particle. For DEM development, Li et al.~\cite{li2010comparison} compared the performances of Verlet buffer and linked-cell approaches in a gravity driven granular collapse simulation. It is reported that appropriate determination of parameters such as search radius (skin + cut-off distance), cell size, and updating interval time step is critical for improving simulation efficiency in Verlet buffer approach.  Although the Verlet buffer mechanism has also been implemented in several DEM codes~\cite{fang2007granular, munjiza2009large}, a uniform skin margin is often adopted for all the particles.
In~\cite{angeles2019assessment}, performance of neighbour search methods (Verlet table and linked cell) and associated computational costs are parametrically evaluated and an evaluation of their suitability for carrying out the DEM/CFD numerical simulations is made. The main outcome of their research showed that the Verlet list has a strong dependency on the skin factor and the value for this parameter equal to the particle radius does not create problems in the identification of particle pairs.
Unfortunately, one noticeable problem is to set the update frequency of the Verlet list according to the uniform skin margin and globally to the maximum velocity leading to stability issues~\cite{li2010comparison, fang2007granular}, considering that particles have possibility of migrating over the skin distance within updating interval. The performance concern of Verlet buffer arises as a result of inhomogeneous flow conditions commonly found in real particle systems, which makes the adoption of a uniform skin margin parameter less computationally efficient. Dynamically determining a local skin margin for each particle according to local flow conditions is suggested in a lot of research, but the optimal determination of the skin margin need to be thoroughly studied. \par

In this research, we proposed a local Verlet buffer approach using a new skin formulation which dynamically expresses the skin margin for each particle according to the neighbourhood flow conditions and based on the particle velocity. This approach enables the inhomogeneity of real particle systems to be taken into account for better computational time efficiency. In this study, the potential stability problem of particles moving over the skin distance in a period of update time is fixed by recording each particle displacement and automatically deciding when the Verlet list is to be rebuilt. We ensured and demonstrated that there is no missed interactions in our current approach and thus our results are identical to using the naive approach.
To assess the efficiency of our proposed formulation, we have implemented the local Verlet buffer approach in our in-house eXtended Discrete Element Method XDEM software~\cite{peters2013extended}. We therefore explored how the skin margin value affects the broad and narrow phase in particular, and the global simulation time in general. 
The main goal of this paper is to propose a broad analysis of our skin formulation implemented in a DEM software like our in-house \textsc{XDEM} toolbox. It points out the advantages of using such formulation and its best-use case but also its drawbacks.

This paper provides a general overview of the XDEM software in background section~\ref{sec:background_xdem} and describes the collision detection method before our current research. The contribution of the article is presented in section~\ref{subsec:verlet_buffer}, which describes the local Verlet buffer approach when building the list of interacting particles and a proof of the method's validity. In section~\ref{sec:param_study}, The skin margin parameter is studied by employing deterministic designs to explore the effect of parametric changes within simulation models. The results and conclusion are discussed in section~\ref{subsec:results}. We finally give a general conclusion of the paper in section~\ref{sec:conclusion}.

%% file: related_work.tex
The Verlet list was first introduced for molecular dynamic simulations by Loup Verlet in his article~\cite{verlet1967computer} back to $1967$. The method is now widely used in DEM simulations and considerably decreases the simulation time. The Verlet list allows to reduce the evaluation of the unnecessary interactions and to keep a neighbour list for several time-steps until a breach. Loup Verlet himself proposed to extend the particles interaction range by a certain skin margin given by:

\begin{equation}
\label{eq:skin_val}
  R_{NL} = R_C + skin ,
\end{equation}

where $R_{NL}$ is interaction range and $R_C$ the cut-off radius. It is then possible to have the exact update interval time step for a given interaction range. Sutmann et al.~\cite{sutmann2006optimization}, Awile et al~.\cite{awile2012fast}, Mattson et al.~\cite{mattson1999near}, and Chialvo et al.~\cite{chialvo1990use} presented different procedure to determine the skin value depending on some parameter as: density, temperature, time step, system size, and molecular geometry.

Chialvo et al.~\cite{chialvo1990use} investigates the effects of the parameters cited above upon the optimum neighbour list radius and update frequency. The theoretical predictions (according to which the optimum neighbour list radius increases with sample size, temperature and time step, and decreases with density) validate the simulation results. The study of the paper is in some ways similar to this paper unlike their study in based on MD simulations while this work focuses on DEM simulations which are very different from MDs in many aspects. In both methods, $N$ or $K$, the update interval time step is not a fixed number but determined by the particles displacement. The difference comes actually from the displacement calculation method: we consider a linear displacement since the last neighbour list update and Chialvo et al. considered in their paper the displacement as the accumulated displacement suffered by the particle since the last neighbour list update.

Sutmann in~\cite{sutmann2006optimization} investigates the performance of neighbour list techniques in MD simulations depending on a variety of parameters, which may be adjusted for maximum efficiency. The model presented allows choosing optimal parameters for the performance of Verlet list and linked-cell lists. The paper targets only Lennard–Jones MD systems.

Awile~\cite{awile2012fast} present a novel adaptive-resolution cell list (AR cell list) algorithm and the associated data structures that provide efficient access to the interaction partners of a particle, independent of the (potentially continuous) spectrum of cut-off radii present in a simulation. They characterise the computational cost of the proposed algorithm for a wide range of resolution spans and particle numbers. Mattson presents a modified method here allowing for reductions in the cell sizes and the number of atoms within the volume encompassing the neighbor cells. The algorithms determine the volume with the minimum number of neighbor cells as a function of cell size and the identities of the neighboring cells. It also evaluates the serial performance as functions of cell size and particle density for comparison with the performance using the conventional cell-linked list method. The two papers by Awile and Mattson target cell linked list method rather than the Verlet list or a combination of the two methods.

LIGGGHTS and LAMMPS software allow setting parameters that affect the building of pairwise neighbour lists. All atom pairs within a neighbour cut-off distance equal to their force cut-off plus the skin distance are stored in the list. The default value for skin depends on the choice of units for the simulation and the inputs.


The methods presented previously and found in the literature mainly focused on MD simulations (some on DEM) where neighbour list updates frequency is usually (not always) fixed beforehand. Our paper in the other hand presents an interaction range ($skin$) formulation based on the velocity of each individual particle. It also focused on an automatic update list technique based on particles displacement avoiding a divergence or crash of the system while reaching optimal performances in DEM simulations.

%% file: background_xdem.tex
The XDEM software is a numerical multi-physics simulation framework~\cite{peters2013extended,samiei2010discrete} supporting parallel processing~\cite{besseron2013unified, checkaraou2018hybrid}, and based on the dynamics of granular material or particles described by the classical DEM~\cite{Cundall79,Allen90}.
It is extended by additional properties such as the thermodynamic state and stress/strain for each particle for more complex simulations in various domains~\cite{peters2019xdem,peters2017Flow10993-31734,Mahmoudi2016d}. As in any DEM code, the particles interaction detection is a major part of XDEM and it uses the linked-cell method to generate the interacting particles list. Firstly, an overview of the XDEM work-flow is provided with the different key parts. Then the collision detection techniques and the different issues making it a major DEM component are presented. Finally, an overview of the linked-cell method and its current implementation in XDEM is given.

\subsection{XDEM flow chart}
\label{subsec:DEM}

\begin{figure}[p]
  \centering
  \captionsetup{justification=centering}
	\includegraphics[width=0.68\linewidth]{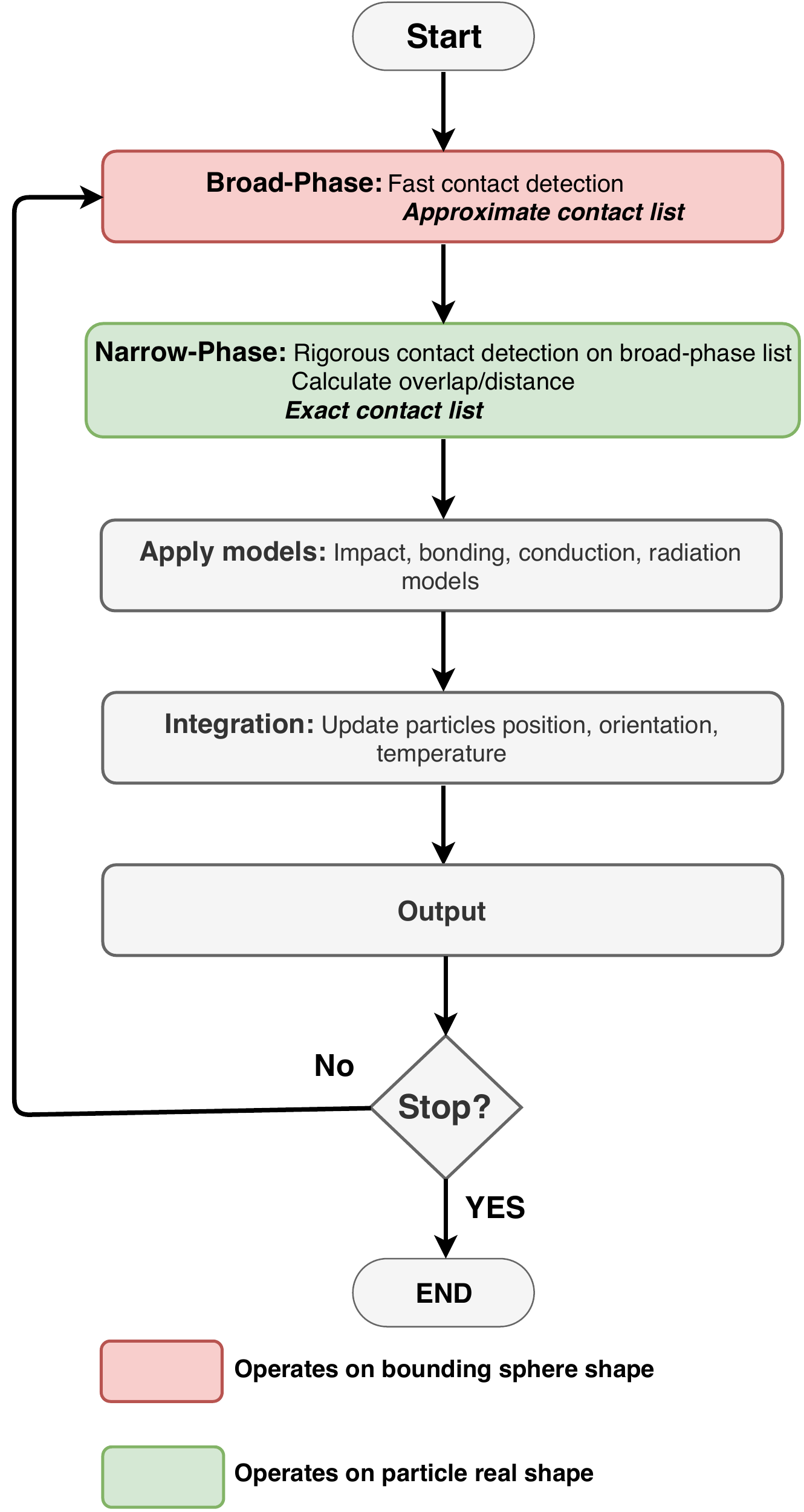}
	\caption{Flow chart of XDEM software showing the main different steps in an iterative simulation.}
	\label{background:fig:xdemworkflow}
\end{figure}

A flow chart as shown in Fig.~\ref{background:fig:xdemworkflow} illustrates the main components of XDEM software for particles dynamics simulation. An iterative time loop is composed of five major phases:
 
\begin{itemize}

	\item \textbf{Broad-phase}: uses a fast but approximate contact detection to build a list of particle pairs that can potentially interact. It should be noted that the pairs of potentially interacting particles are stored in a unique list. During this phase, the particles are represented by a bounding volume shape. It builds the list of interacting particles pairs by dividing the domain in cell with uniform size using the linked-cell method. The broad-phase represents around $65\%$ of the total computational time;
	
	\item \textbf{Narrow-phase}: performs a rigorous contact detection of each pair of particles in the broad-phase list using the actual shape of the particle and calculates the pairwise collision parameters such as overlap,  contact location and direction. The XDEM software supports complex shapes by using the sub-shapes techniques (a shape is composed by many simple shapes as spheres) and super-quadratics. The narrow-phase represents around $15\%$ of the total computational time;
	
	\item \textbf{Apply physical models}: based on the collision parameters and collision history information, calculate all interaction forces by applying corresponding physical models such as normal and tangential contact models, rolling models, and cohesive models and so on;
	
	\item \textbf{Integration}: updates the particle location, velocity, rotational velocity and orientation information by numerically integrating Newton's second law with various numerical algorithms such as leapfrog and velocity-Verlet schemes;
	
	
\end{itemize} 

\par

\subsection{Collision detection in XDEM}
\label{subsec:collision_detection}

The contact detection being carried out on a large number of particles, it is split into two phases in order to reduces the computational complexity: a first, fast and approximate phase called the \textbf{broad-phase} and an accurate second phase called \textbf{narrow-phase} as indicated in Fig.~\ref{background:fig:xdemworkflow} and already mentioned in Section~\ref{subsec:DEM}. \par 

\begin{figure}[htbp]
  \centering
  \captionsetup{justification=centering}
  \includegraphics[width=\linewidth]{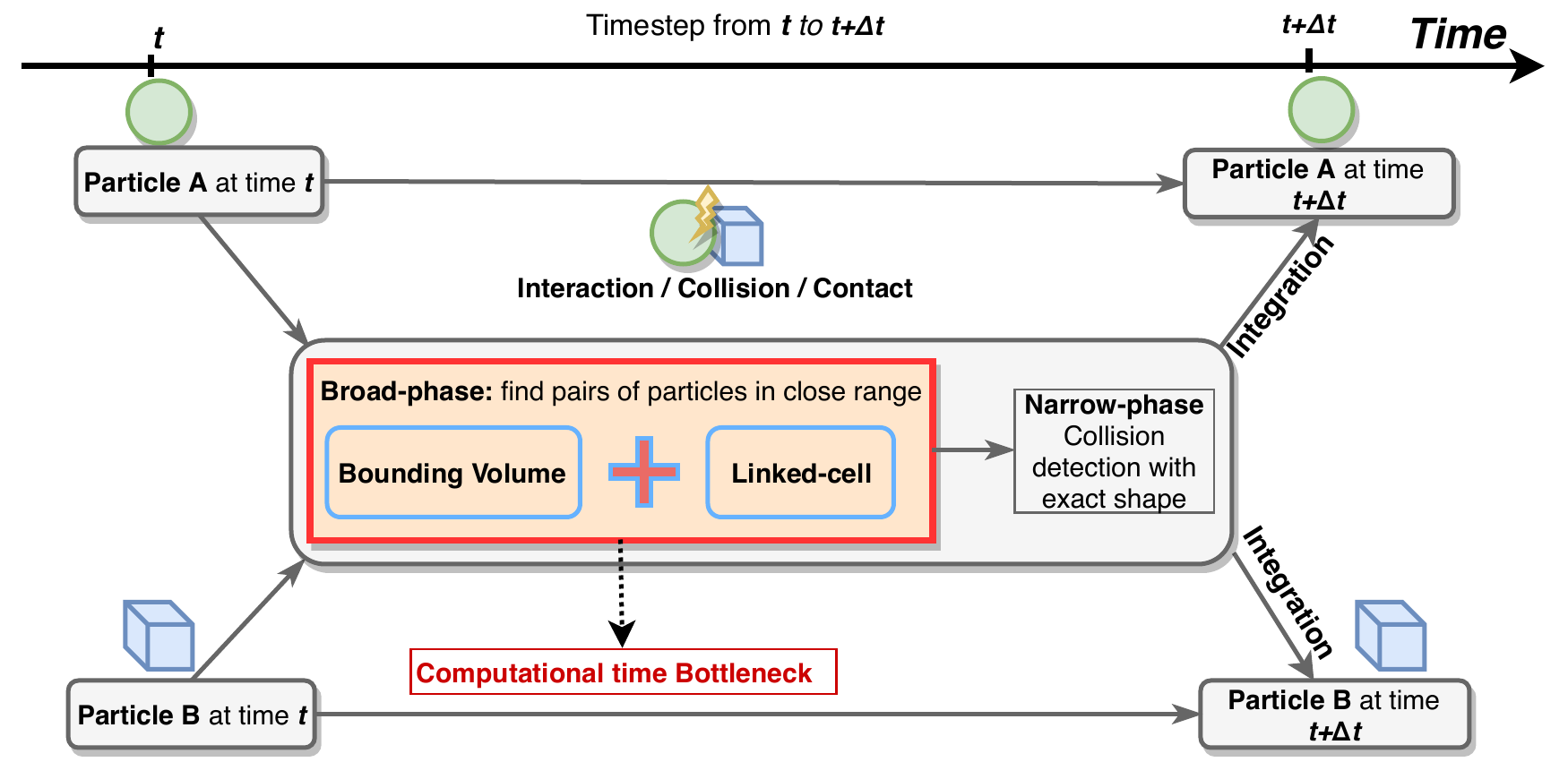}
  \caption{Collision detection (broad-phase and narrow-phase) process workload in XDEM. The broad-phase is the main computational time consumer.
  }
  \label{background:fig:coldetect}
\end{figure}

Fig.~\ref{background:fig:coldetect} illustrates the collision detection process for two colliding particles of any shapes in XDEM. A bounding volume enclosing any type of particle entity replacing the particle real shape is used to achieve a rapid broad-phase detection. In \textsc{XDEM} software, the broad-phase is carried on using bounding spheres, slightly increasing the memory usage (the BS requires additional data) but greatly improves/reduces data accesses from the CPU. With the BSs, the distance computation becomes less computationally expensive.

Realising that two particles far away from each other have very little chance to generate any interaction, the neighbour particles detection is often limited to a certain distance. The Algorithm~\ref{algo:linked_cell} describes the linked-cell technique used in XDEM to spatially limit the contact detection process of a pair of particles.

\bigskip 

\begin{algorithm}[htbp]
\SetAlgoLined
 {Uniform decomposition of the domain in cells}\;
 \For{all $C_a$ in cell list}{
  \For{$C_b$ in the immediate neighbour of $C_a$}{
    \tcp{Make sure to check each pair of cells only once}
    \If{index($C_b$) $<$ index($C_a$)}{
       \For{each $P_a$ among all particles in cell $C_a$}{
         \For{each $P_b$ amoung all particles in cell $C_b$}{
           \tcp{Check if the bounding spheres of $P_a$ and $P_b$ intersect}
           \If{$\vert \vert X_a-X_b\vert \vert \leq r_a+r_b$}{
	       $List \gets ({P_a,P_b})$\;
	       }
         }
       }
    }
  }
 }
 \caption{Linked-cell algorithm}
 \label{algo:linked_cell}
\end{algorithm}

\bigskip

\begin{figure}[htbp]
  \centering
  \captionsetup{justification=centering}
  \includegraphics[width=0.45\linewidth]{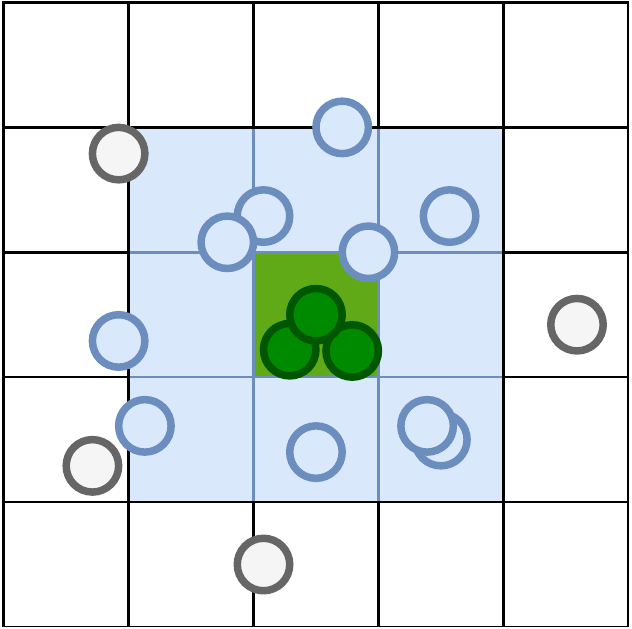}
  \caption{Illustration of the cell linked method. For the particles in green cell, a collision is checked with particles in the same cell (green particles), and also within the immediate neighbour cells (blue).}
  \label{cell_linked_list}
\end{figure}

The linked-cell approach is utilised to perform the neighbour list construction, which guarantees the time complexity is linear in the number of particles in the system.  As illustrated in Fig.~\ref{cell_linked_list}, the pairwise interactions for a single particle is limited with all particles within the same cell (green) and in the immediate or adjacent neighbouring cells (blue). The cell size is uniform and must not be smaller than the maximum bounding sphere size of all particle entities.

The narrow-phase using the particles real shapes is performed on the broad list of interacting pairs of particle returned by the broad-phase. The time complexity for narrow-phase collision detection is largely determined by the particle shape. For spherical particles, the narrow-phase detection is simply checked following identity:

\begin{equation}
   \delta = r_i+r_j-\vert \vert X_i-X_j\vert \vert
   \label{eq:delta}
\end{equation}
\noindent where, $r_i$ represents the radius of particle $i$, and $X_i$ is the center coordinates of particle $i$. If the overlap $\delta$ is positive, the two particles collide and vice versa. For other particle shapes such as superquadratics, the collision detection becomes more complex and usually an optimization problem need to be solved~\cite{williams1992superquadrics}.

%% file: verlet_buffer.tex
The Verlet buffer, unlike the conventional Verlet list, does not build a neighbourhood list for each particle but rather a global list of interacting particle pairs also called the Verlet list. In both methods, the particle cut-off radius (bouding sphere in DEM) is surrounded by a skin margin~\cite{Allen90, allen2017computer}, to give a larger neighbourhood. Another difference lies in the ability to work with any broad-phase algorithm. 

In our approach, we extend the bounding spheres used by the broad-phase to perform its approximate collision detection.
The extension range of the bounding spheres is called \textbf{skin}, and it will increase the number of potential interactions found by the broad-phase by considering pair of particles that are located further away from each others. On one side, this will make the broad-phase and the narrow-phase costlier to evaluate, but on the other side, the broad-phase does not have to be executed at every time step anymore. By considering a larger surrounding in the broad-phase collision detection, the list of potential interactions now includes interactions that could happen in the next time steps. 

In the local buffer method, the skin margin used to extend the bounding volumes is unique to every particle and is computed according to their local flow conditions. Additionally, we propose a condition that allows to check that the result of the previous broad-phase (i.e. the list of potential interactions) is still correct~\cite{noske2004efficient}. When this condition is broken, we can force the execution of a new broad-phase.
In any case, the narrow-phase is always executed on this approximated list of interactions, and that, guarantees that the results will be strictly identical with the case of having the broad-phase always executed.

\begin{figure}[htbp]
  \centering
  \captionsetup{justification=centering}
  \begin{subfigure}[b]{0.45\textwidth}
    \includegraphics[width=\linewidth]{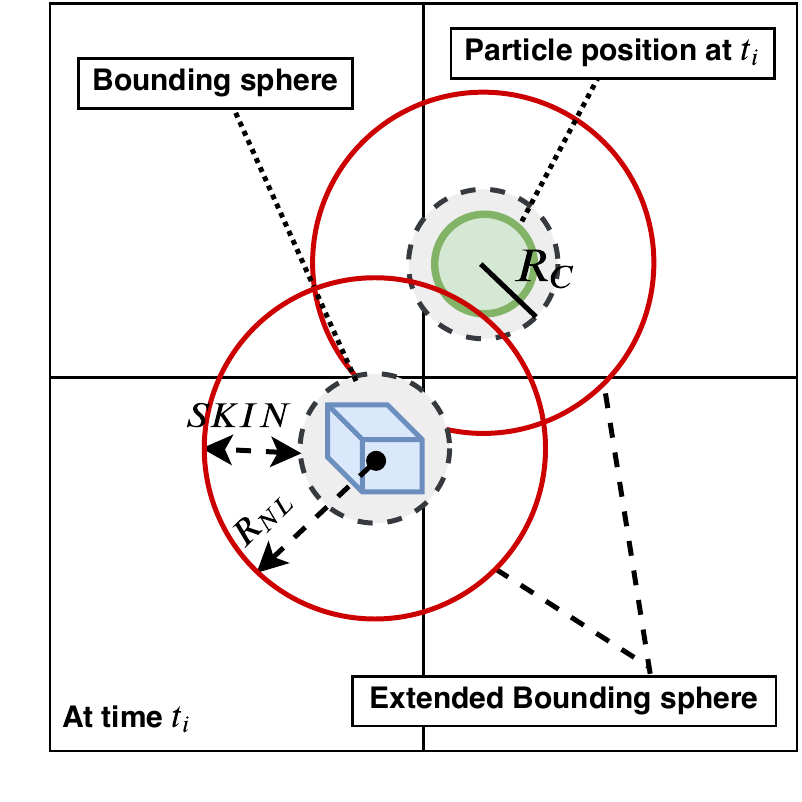}
    \caption{Extension of the interaction range by surrounding the cut-off radius by a skin margin.}
    \label{fig:first}
  \end{subfigure}\\
  \bigskip
  \begin{subfigure}[b]{0.45\textwidth}
    \includegraphics[width=\linewidth]{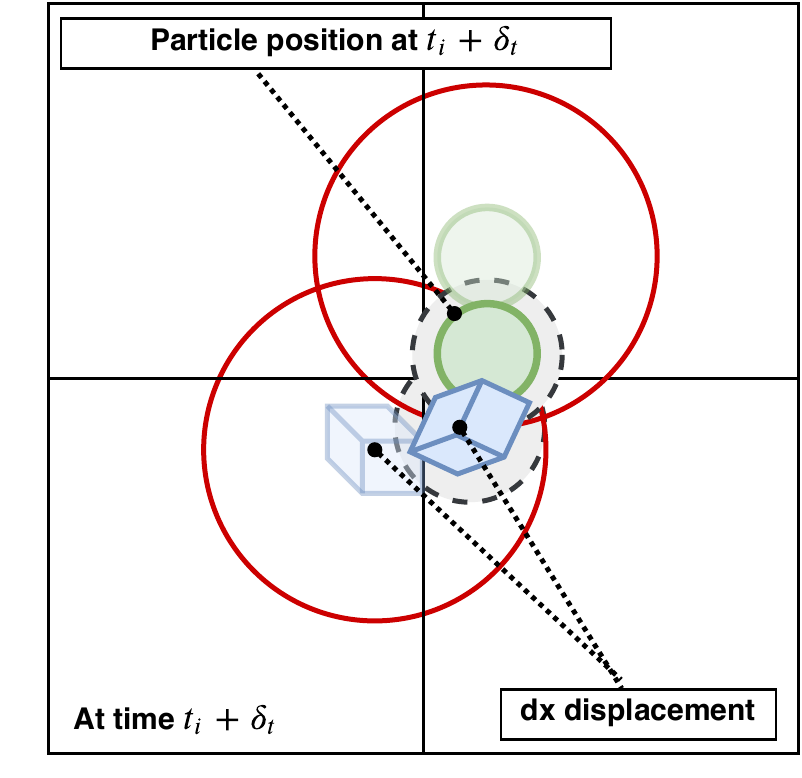}
    \caption{Collision in extended interaction range when the two particles are still in the Verlet list. The narrow-phase is applied to check the actual collision.}
    \label{fig:second}
  \end{subfigure}
  \hfill
  \begin{subfigure}[b]{0.45\textwidth}
    \includegraphics[width=\linewidth]{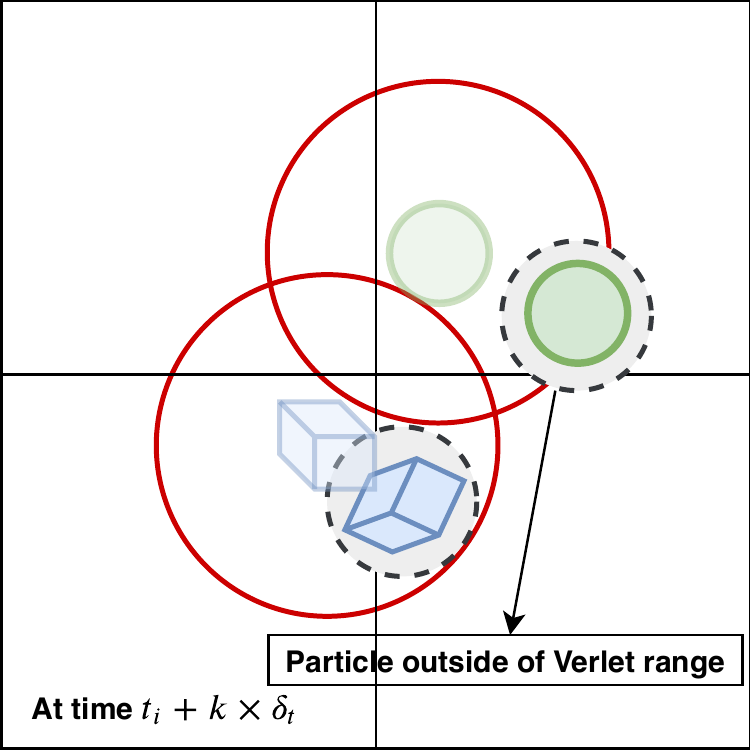}
    \caption{New broad-phase required, the green particle have moved for more than a skin distance(and could have moved in another particle's neighborhood).}
    \label{fig:third}
  \end{subfigure}
  \caption{Initial configuration and update of the Verlet list.}
  \label{fig:first_second}
\end{figure}

The application of Verlet buffer in particle collision detection process is illustrated in Fig.~\ref{fig:first_second}. On the example of Fig.~\ref{fig:first}, the two particles, with different shapes, have their bounding spheres extended by a skin margin often called the Verlet skin. This phase extends each particle neighbourhood and includes the pair of particles in the Verlet list, which without the extension, would have not been considered as potentially interacting. 
The Fig.~\ref{fig:second} shows that after a couple of steps, the Verlet list does not need to be updated.
The particles, or more precisely the bounding sphere of each particle, did not leave their respective extended bounding spheres
that where use to perform the broad-phase collision detection initially.
That shows how an extension of a skin margin in the particles neighbourhood includes the pair of particles in the Verlet list 
and catches an active collision that happens a few steps later. 
Finally, Fig.~\ref{fig:third} illustrates the case where one of the particles moves out of its skin margin. This means that the Verlet list is not valid anymore and the broad-phase must be executed again.

\begin{figure}[htbp]
  \centering
  \captionsetup{justification=centering}
	\includegraphics[width=0.65\linewidth]{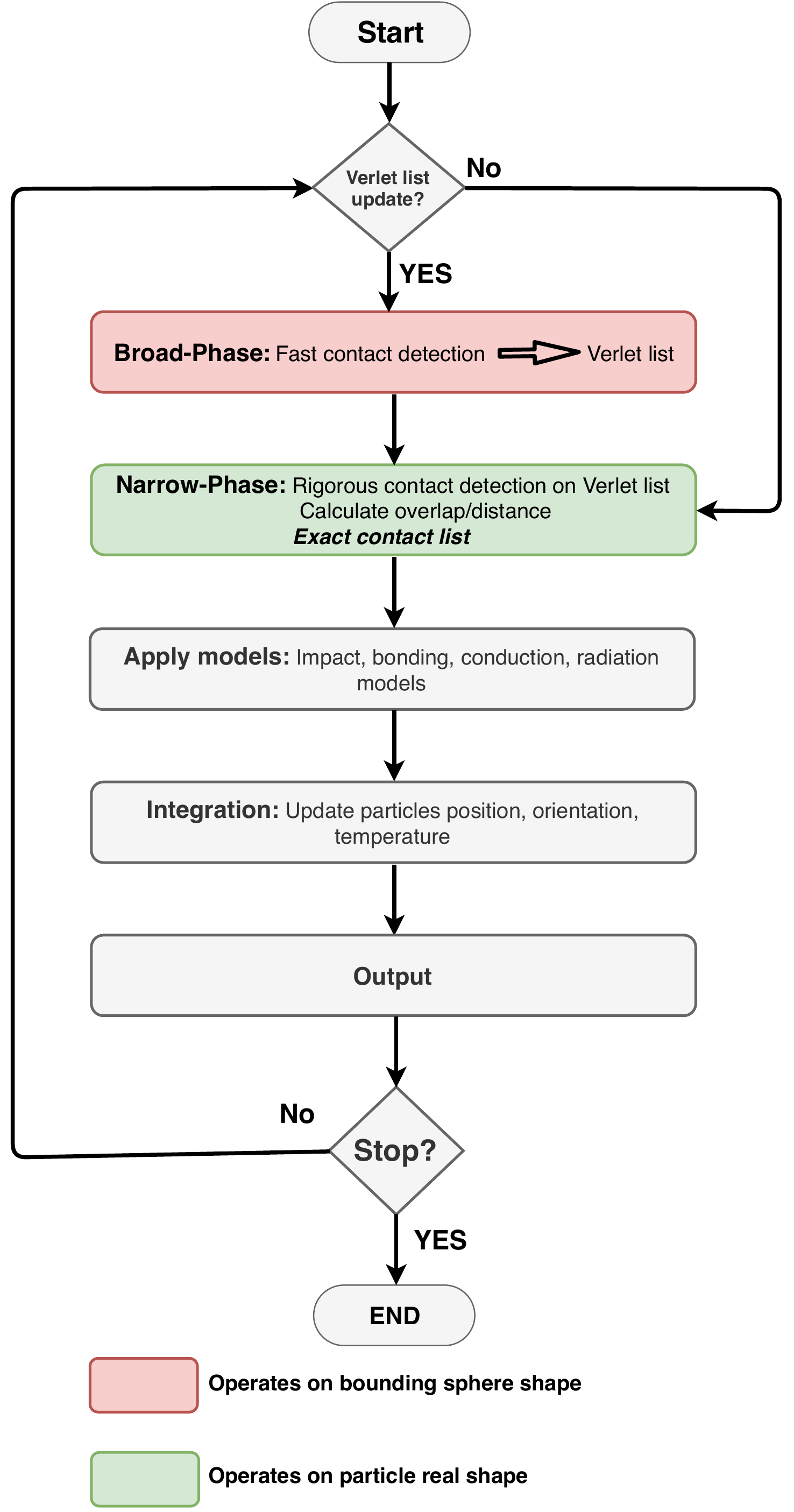}
	\caption{New flow chart of XDEM software. A construction of the Verlet list is added. If the conditions are satisfied, the list is kept and the broad-phase is skipped and the simulation continues directly to the narrow-phase.}
	\label{fig:xdemw_flow_chart}
\end{figure}

The overall procedure for constructing the Verlet buffer list in XDEM is shown in Fig.~\ref{fig:xdemw_flow_chart}

During a broad-phase collision detection, for a particle $i$, every particle $j$ in its neighbour cell is checked, as described in linked-cell method, to determine whether the pair $(i,j)$ should be included in the Verlet list or not. Particle $i$ and $j$ constitute a pair in the Verlet list, if they satisfy the condition that the bounding spheres for the two particles overlap with each other. The radius of the bounding volume for particle $i$, called the neighbourhood list radius is calculated as:

\begin{equation}
    R_{NL,i} = R_{C,i}+skin_i,
    \label{eq:bounding_sphere}
\end{equation}

\noindent where, $R_{C,i}$ is the interaction cut-off radius determined by applied physical models, and $skin_i$ is the local Verlet skin distance, the value of which depends on the local flow properties.
The construction of Verlet list process loops over all the pairs of particles in the linked-cell neighbor list for completion. In the next following steps, the displacement of each particle starting since the last Verlet list build is examined against a local threshold. The Verlet list remains unchanged and the broad-phase collision detection is skipped until a violation occurs. The following talks about how to determine the local skin distance and a scheme for automatically updating Verlet list in this approach.

\subsection{The local skin parameter}
\label{subsubsec:skin}
To choose the skin distance parameter, we extend the equation proposed by Loup Verlet in 1967~\cite{verlet1967computer} for MD
and we propose to use the formula:

\begin{equation}
skin_p = K \times v_p . \Delta t
\label{eq:skin_formulus1}
\end{equation}

\noindent where, $v_p$ is the particle velocity in the system, $\Delta t$ is the time step in the simulation, and $K$ is the prescribed number of skipped steps in the broad-phase collision detection. This research started with Eq.~\ref{eq:skin_formulus1} to determine the local skin parameter for each particle by replacing bulk velocity $v$ with particle local velocity $v_p$. 

\subsection{Automatic update and validity of the Verlet list}
\label{subsubsec:scheme}

In this section, we detail a condition that allows to determine if the Verlet list, computed during a previous time step, is still valid~\citep{grindon2004large}.

\begin{definition}
The Verlet list $V_L$ at time $t$ is correct if 
$$\| \vec{AB} \|_{t} \leqslant R_{A} + R_{B} \implies \left\{ A,B \right\} \in V_L$$
\label{def:verletlist}
\end{definition}

We use the following condition as a way to determine if the Verlet list at time $t$ is still valid

\begin{equation}
    \forall \ \mathrm{particle} \ p, \ \Delta x_p \leqslant skin_p
    \label{eq:update}
\end{equation}

where $\Delta x_p$ is the particle displacement of the particle since the last broad-phase.
If this condition is violated, it means that a new broad-phase must be executed to update the Verlet list.

\begin{definition}
We define our \textbf{automatic Verlet update} scheme by:
\begin{itemize}
    \item re-using the previous Verlet list while condition~\ref{eq:update} is still valid;
    \item re-executing a new broad-phase to update the Verlet list otherwise.
\end{itemize}
\label{def:autoverlet}
\end{definition}

We will now prove that our proposed automatic Verlet update scheme always returns the correct results.

\begin{theorem}
	The automatic Verlet update scheme defined at~\ref{def:autoverlet} ensures that all pairs of particles that can possibly collide are in the Verlet list.
\label{theo:update_verlet}
\end{theorem}
	
The claim in theorem~\ref{theo:update_verlet} avoids the potential simulation accuracy and stability issues that are founded in the fixed update interval scheme as reported in~\cite{chialvo1990use}. The proof of the theorem is provided in following step:

\begin{proof}

We assume that the initial broad-phase executed at time $t$, performed on the extended bounding spheres, is correct and returned the correct Verlet list, and thus by definition~\ref{def:verletlist}, we have:

\begin{equation}
\label{eq:broad_phase}
\| \vec{AB} \|_{t} \leqslant R_{NLA} + R_{NLB} \implies \left\{ A,B \right\} \in V_L,
\end{equation}

where $R_{NLA}$ and $R_{NLB}$ are the respective extended interaction ranges of particle A and particle B.

We want to show that, at any time $t^{'} = t + \Delta t$, the Verlet list is valid.
According to the definition~\ref{def:verletlist}, that means that considering $\left\{ A,B \right\}$ a pair of particles, we need to prove the hypothesis H0, that if A and B collide at time $t'$, then $\left\{ A,B \right\}$ is in the Verlet list $V_L$:

\begin{equation}
\label{eq:hypothesis}
\| \vec{AB} \|_{t'} \leqslant R_{CA} + R_{CB} \implies \left\{ A,B \right\} \in V_L,  \quad \quad (H0)
\end{equation}

where $R_{CA}$ and $R_{CB}$ are the respective cutoff distance of particle A and B.

It exist two possibilities:

\begin{enumerate}

  \item The condition~\ref{eq:update} is no more valid, i.e.:
    \begin{equation}
    \label{eq:non_valid_cond}
      \exists A \: | \: \| \vec{AA'} \|_{t'} > skin_A,
    \end{equation}
    where $A$ is a particle at a given position at time $t$, $A'$ the same particle in a new position at $t'$ time. 
    In that case, a new broad-phase, on the extended bounding spheres, has to be executed at time $t'$, and a new correct Verlet list is generated. 
    So if we consider two colliding particles A and B at time $t'$, then we have

    \begin{equation}
    \begin{aligned}
    \label{eq:non_valid_cond2}
      \: \| \vec{AB} \|_{t'} & \leqslant R_{CA} + R_{CB}  & & \text{\footnotesize because A and B are interacting} \\
                             & \leqslant R_{CA} + skin_A + R_{CB} + skin_B \\
                             & \leqslant R_{NLA} + R_{NLB} \\
    \end{aligned}
    \end{equation}
    and because the newly generated Verlet list is correct, then we have $\left\{ A,B \right\} \in V_L$.
    The hypothesis H0 is verified.

    \item The condition~\ref{eq:update} is still valid, i.e.:

    \begin{equation}
    \label{eq:valid_cond}
      \forall A, \: \| \vec{AA'} \| \leqslant skin_A
    \end{equation}

    where $\| \vec{AA'} \|_{t'}$ is the distance covered by particle $A$ from time $t$ to $t'$.

    So if we consider two colliding particles A and B at time $t'$, then we have
    \begin{flalign*}
      & \| \vec{AB} \|_{t'} \leqslant R_{CA} + R_{CB} \\
      & \implies  \: \| \vec{AB} \|_{t'} + \| \vec{AA'} \| \leqslant R_{CA} + R_{CB} + skin_A & \text{\footnotesize by adding Eq.~\ref{eq:valid_cond} for particle A}  \\
      & \implies \: \| \vec{AB} \|_{t'} + \| \vec{AA'} \| + \| \vec{BB'} \| \leqslant R_{CA} + \ R_{CB} + skin_A + skin_B & \text{\footnotesize by adding Eq.~\ref{eq:valid_cond} for particle B} \\
      & \implies \: \| \vec{AB} \|_{t'} + \| \vec{AA'} \| + \| \vec{BB'} \| \leqslant  R_{NLA} + R_{NLB} \\
      & \implies \: \| \vec{A'B'} \| + \| \vec{AA'} \| + \| \vec{BB'} \| \leqslant  R_{NLA} + R_{NLB} & \text{\footnotesize as $\| \vec{AB} \|_{t'} = \| \vec{A'B'} \|$} \\
    \end{flalign*}

    Additionally, from the triangle inequality, we have
    $$\| \vec{AB} \| \leqslant \| \vec{AA'} \| + \| \vec{A'B} \|  \leqslant \| \vec{AA'} \| + \| \vec{A'B'} \| + \| \vec{B'B} \|$$

    and which finally gives us

    \begin{equation*}
    \begin{aligned}
      \| \vec{AB} \|_{t'} \leqslant R_{CA} + R_{CB} & \implies \| \vec{AB} \|_{t'} \leqslant R_{NLA} + R_{NLB} \\
                                                    & \implies \left\{ A,B \right\} \in V_L & \text{\footnotesize because of Eq.~\ref{eq:broad_phase}} \\
    \end{aligned}
    \end{equation*}
    The hypothesis H0 is verified.

\end{enumerate}

\end{proof}

%% file: performance_eval.tex
We carried out extensive numerical experiments to assess the performance of our proposed approach.

\subsection{Methodology}

In section~\ref{subsec:verlet_buffer}, we presented our first model to establish the skin margin value of a particle:

\begin{equation}
skin_p = K \times v_p . \Delta t
\label{eq:skin_formulus}
\end{equation}

where $v_p$ is the particle velocity in the system, $\Delta t$ is the time step in the simulation and $K$ is the prescribed number of skipped steps in the broad-phase collision detection.

Dynamic determination of the skin margin allows for particles from different flow velocity adopting distinct skin margins, but with the same $K$ number of time steps between two consecutive updates of the Verlet list. What is the optimum $K$ value giving the best computational efficiency? To answer this question, we performed a parameter study on the skin margin by varying the value of $K$. It varies from $0$ to $5000$ in the current study that was conducted on five different real test-cases with different purposes and flow velocity. 

In the following XDEM simulations, we used the Velocity Verlet integration scheme and the Linear Spring Dashpot III contact model (only the static friction force is taken into account in the classical Linear Spring Dashpot model). All the five test-cases have been simulated for at least $5000$ time steps. The Verlet buffer method is coupled with the linked-cell method with the constraint on the neighbour list range to be smaller than the cell size. This gives an upper-bound to the skin value $R_{NL} <= \min_{x, y, z} L_C$.
This means that in practice, the skin will be set for each particle independently to

\begin{equation}
R_{NL,p} = min( R_{C,p} + skin_p, \min_{x, y, z} L_C  )
\label{eq:skin_formulus_mincell}
\end{equation}

\subsection{Test-cases}
\label{subsec:testcases}

The following real-world example serves as concrete benchmarks for the evaluation of our implementation.

\begin{itemize}
\item \textbf{Hopper Discharge}
The \textit{Hopper Discharge} presented in Fig.~\ref{fig:hopper_setup} is a test case used with $125k$, $250k$ and $500k$ particles.
The simulation works as follows: the selected number of particles (thus up to half a million) with different diameter are dropped off in a silo. Then the notch at the bottom is opened, letting all particles fall down into a chute.
In this case study, the workload moves from the top portion of the domain downwards.
Since the lower part of the silo is narrowed, the workload is focused on the center region of the domain.

\begin{figure}[htbp]
  \centering
  \captionsetup{justification=centering}
  \includegraphics[width=\linewidth]{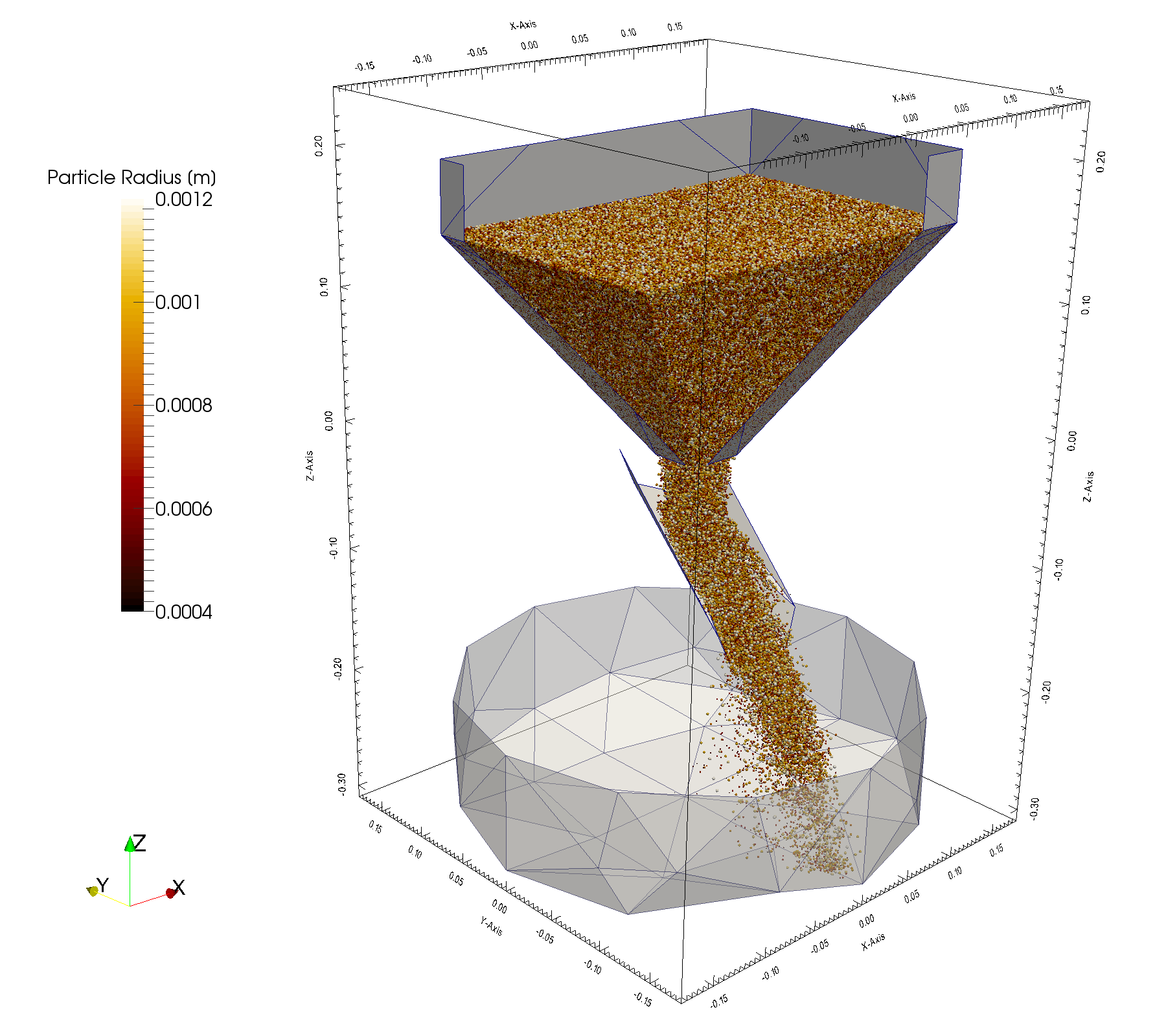}
  \caption{The test case used for the performance evaluation simulates the hopper discharge of 125k, 250k and half a million particles. It shows any overview of the set-up with the particles coloured according to their size.}
  \label{fig:hopper_setup}
\end{figure}

\item \textbf{Granular flows on vibrating rough inclined planes}

The test case on Fig.~\ref{fig:incline_setup} simulate a granular (spherical particles) flowing down on a roughed and incline plane. The particles are coloured according to their velocity. In this test case, a silo is filled with particles of uniform size that are dropped off on an incline plane. The latter has a roughed surface composed by many bigger particles laterally vibrating. In the free flow of particles, the ones on the top of bed present a higher velocity because being the least in contact with the rough surface and therefore undergoing the least lateral vibrations.

\begin{figure}[htbp]
  \centering
  \captionsetup{justification=centering}
  \includegraphics[width=\linewidth]{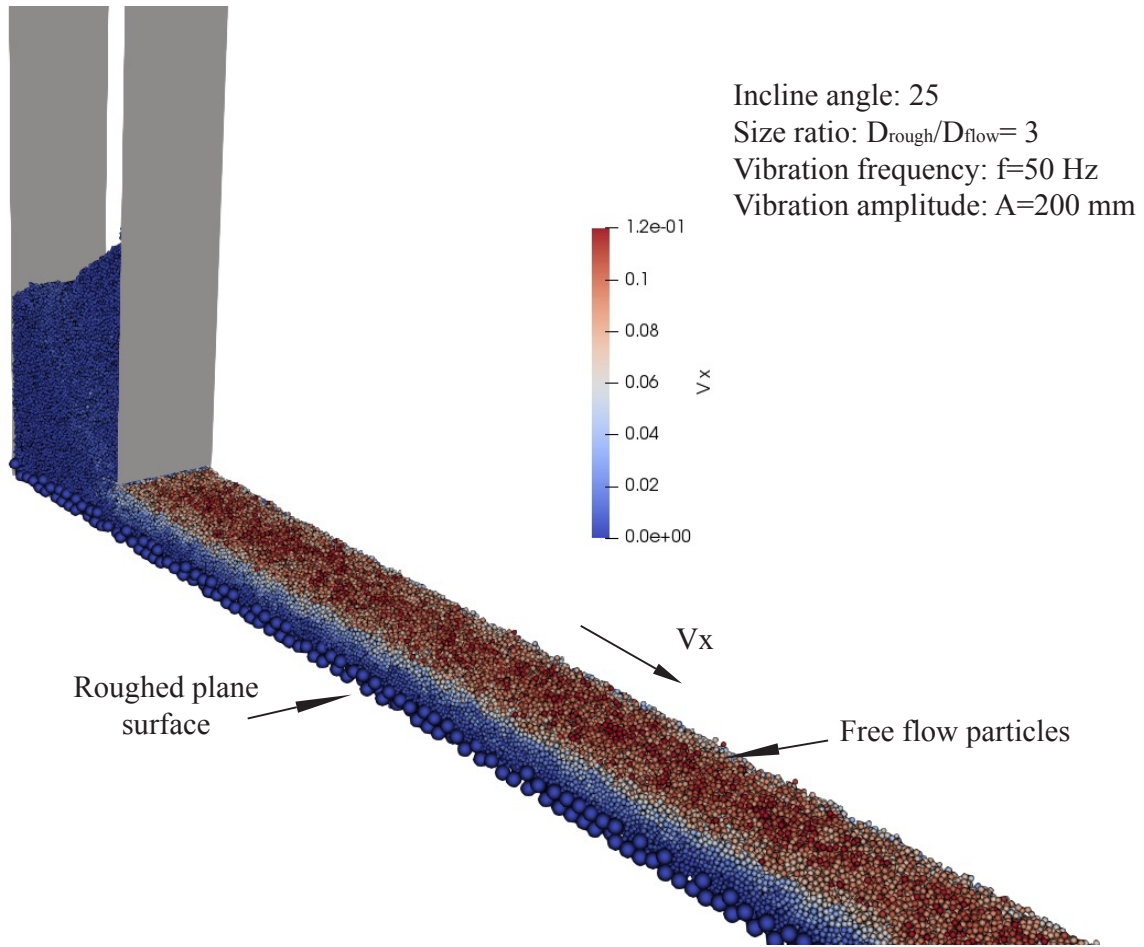}
  \caption{Granular flows on roughed inclined plane. The rough plane has particles vibrating at $50Hz$ frequency with $200mm$ amplitude. The free flow particles are coloured according to their velocity}
  \label{fig:incline_setup}
\end{figure}

\item \textbf{Avalanche}

Fig.~\ref{fig:avalanche} describes a simulation of an avalanche debut at the top of a habitable valley. Particles located upstream of the valley, descend throughout the valley with increased speed due to the inclined surface. The goal is to predict the path and the rate at which the avalanche will reach the bottom of the valley at the housing level. The case, a cohesive model,  uses the Elastic-plastic spring-dashpot rolling model.

\begin{figure}[htbp]
  \centering
  \captionsetup{justification=centering}
  \includegraphics[width=\linewidth]{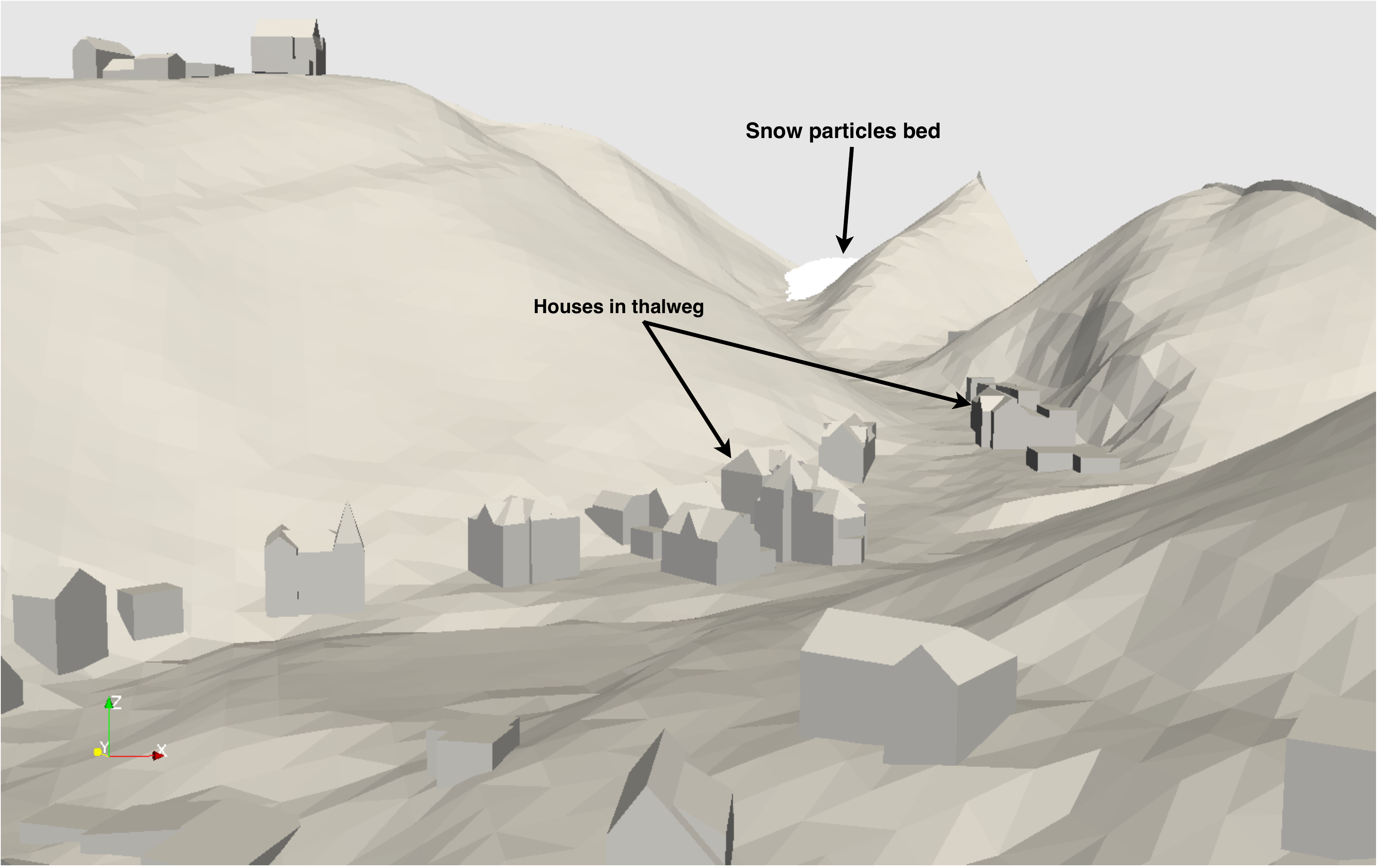}
  \caption{Simulation of an avalanche at the top of a habitable valley. The particles bed represents a cohesive snow model.}
  \label{fig:avalanche}
\end{figure}

\item \textbf{Biomass combustion}

Fig.~\ref{fig:alesio} shows a simulation of a combustion chamber of a 16 MW geothermal steam super-heater, which is part of the Enel Green Power "Cornia 2" power plant~\cite{wiki:xxx}.
The test case relies on a hybrid four-way coupling between the Discrete Element Method (DEM) and the Computational Fluid Dynamics (CFD). In this approach, particles are treated as discrete elements that are coupled by heat, mass, and momentum transfer to the surrounding gas as a continuous phase. For individual wood particles, besides the equations of motion, the differential conservation equations for mass, heat, and momentum are solved, which describe the thermodynamic state during thermal conversion.

\begin{figure}[htbp]
  \centering
  \captionsetup{justification=centering}

  \includegraphics[width=0.8\textwidth]{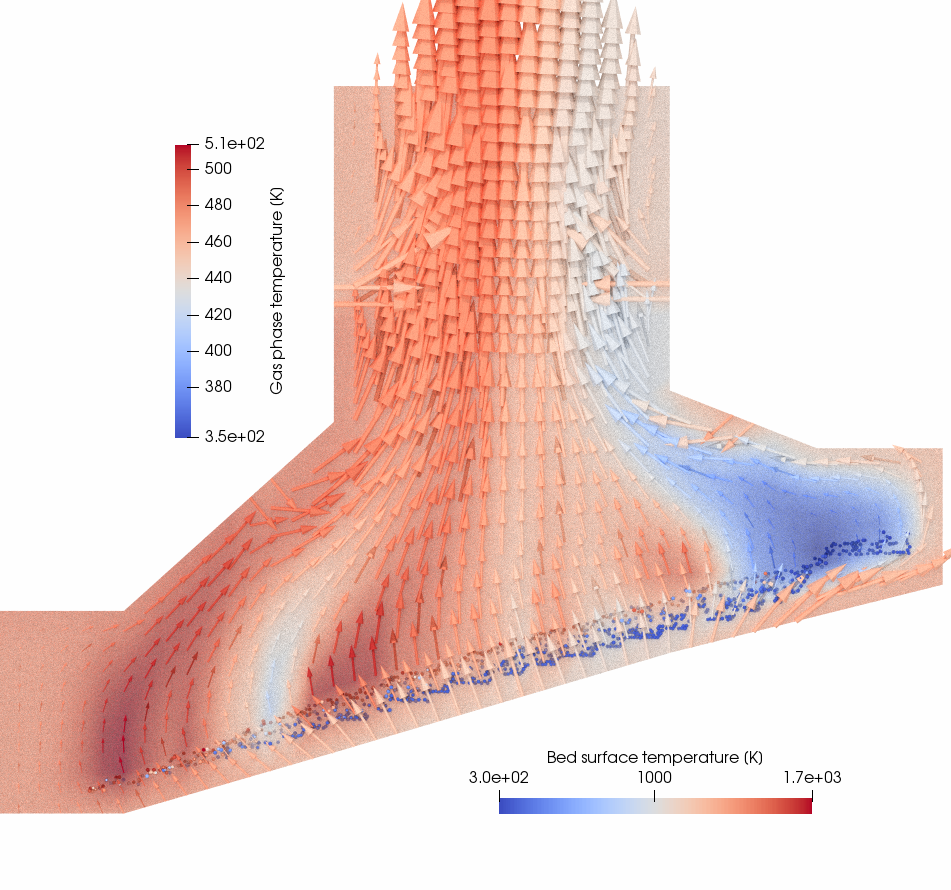}
  \caption{Furnace of the combustion of biomass. The particle bed are arranged on four (4) moving grates. The bed is heated up in the combustion chamber by inlets located below the grates. The particles are coloured according to their surface temperature.}
  \label{fig:alesio}
\end{figure}

\item \textbf{Powder levelling for Selective Laser Melting in Additive Manufacturing}

Additive manufacturing and specifically metal selective laser melting (SLM) processes are rapidly being industrialized~\cite{donoso2018exploring}. The case showed in Fig.~\ref{fig:alvaro} simulates a Powder levelling for Selective Laser Melting (SLM) in Additive Manufacturing (AM) process. In this test case, an advance discrete-continuous concept is used to address the physical phenomena involved during laser powder bed fusion. The concept treats the powder as a set of particles by XDEM, predicting the thermodynamic state and phase change of each particle. The fluid surrounding is solved with multiphase CFD techniques to determine the momentum, heat, gas and liquid transfer.

\begin{figure}[htbp]
  \centering
  \captionsetup{justification=centering}
  \includegraphics[width=\linewidth]{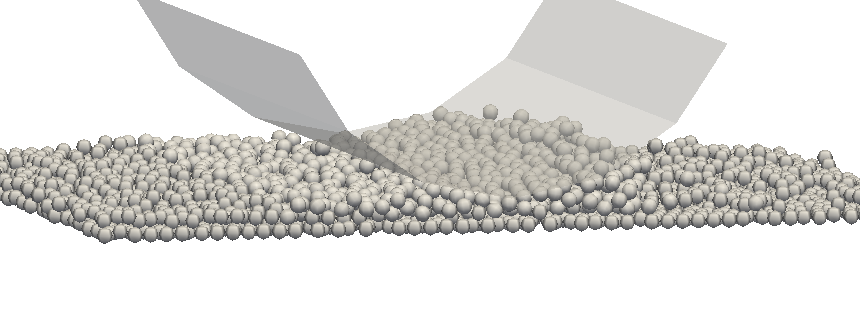}
  \caption{Powder levelling for Selective Laser Melting.}
  \label{fig:alvaro}
\end{figure}

\end{itemize}

\subsection{Experimental settings}

The experiments were carried out using the \texttt{Iris} cluster of the University of Luxembourg~\cite{VBCG_HPCS14} which provides 168 computing nodes for a total of 4704 cores.
The nodes used in this study feature a total a 128 GB of memory and have two Intel Xeon E5-2680 v4 processors running at 2.4 GHz, that is to say, a total of 28 cores per node.
The nodes are connected through a fast low-latency EDR InfiniBand (100Gb/s) network organized over a fat tree topology.

We used version \texttt{67f029de} of XDEM software, compiled with GCC Compiler 6.4.0. 
To ensure the stability of the measurements, the nodes were reserved for exclusive access. Additionally, each performance value reported in this section is the average of at least a hundred measurements. The standard deviation showed no significant variation in the results, and hence is not shown of the following plots.

\subsection{Results}
\label{subsec:results}

\begin{figure*}[htbp]
  \centering
  \captionsetup{justification=centering}
  \includegraphics[width=1.\linewidth]{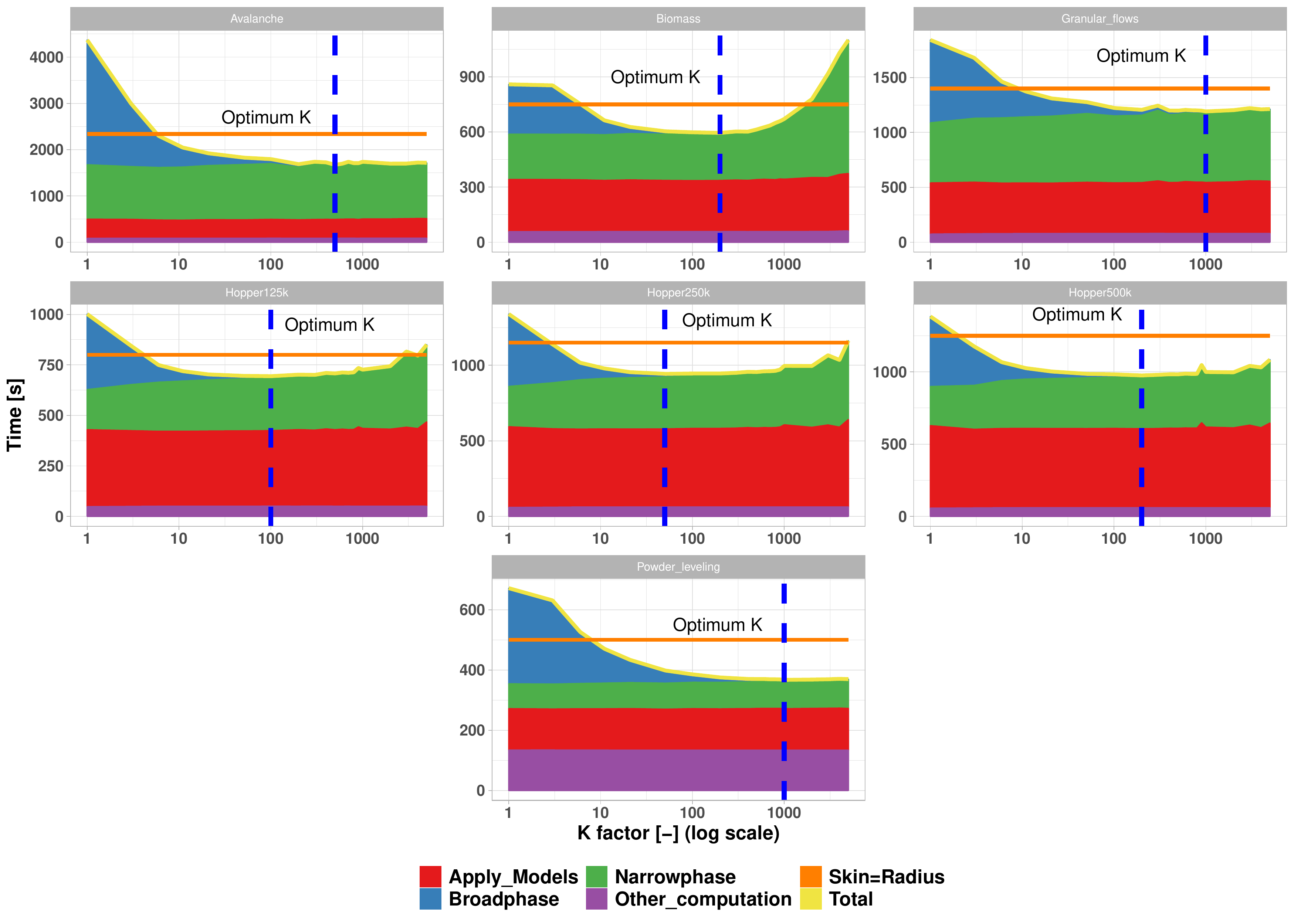}
  \caption{Dependence of broad-phase, narrow-phase and interactions models on skin $K$ factor. The vertical blue dashed lines show the optimum $K$ for each simulations corresponding to the lowest overall simulation time. The orange horizontal line represents the simulation time for a constant skin equal to the particles radius. The skin distance is capped by the cell size in all simulations.}
  \label{fig:geom_area}
\end{figure*}

The different simulations that have been conducted in this study are intended to analyse and interpret the impact of the skin margin upon the performances of the automatic update algorithm presented in subsection~\ref{subsec:verlet_buffer}. In this section we take a look on how the $skin = Kv\Delta t$ model affects the test cases computational times. The simulation time of the broad-phase, narrow-phase and apply models are illustrated in Fig.~\ref{fig:geom_area} as a function of $K$ factor. We also compare our model of $skin = Kv\Delta t$ (skin is different for each particle) with the popular approach where the skin is uniform and equal to the particle radius~\cite{angeles2019assessment} (all cases are monodisperse and therefore the skin is identical for all the particles).

It follows from Fig.~\ref{fig:geom_area} that:

\begin{itemize}

    \item The broad-phase simulation time decreases with the $K$ value and therefore with the $skin$. For the Avalanche case, $BP\footnote{Broad-phase simulation time} = 3172.758s$ without the Verlet buffer method and $BP = 42.418s$ for $K=400$, representing a $98.66\%$ of time improvement.
    
    Increasing the $K$ factor decreases the overall broad-phase time, but does increase a single broad-phase time due to the extend of the neighbourhood (more pair of particles in the Verlet list). But this goes hand in hand with a decrease of the number of performed broad-phase, which decreases the overall time.
    
    \item The narrow-phase simulation time increase with the $K$ value and thus with $skin$. For the Biomass case, $NP\footnote{Narrow-phase simulation time} = 288.155s$ without the Verlet buffer method and $NP = 323.626s$ for $K=1000$. It is a $10.96 \%$ increase in the narrow-phase time.  
    
    The rise is due to the enlargement of the Verlet list during an increase of $K$, on which the narrow-phase performs an exact collision detection.
    
    \item When K increases, the simulation time decreases to a low peak before starting to increase for almost all the case, specially for the Biomass case. Without using the Verlet buffer method, simulation time equals to $962.901s$ and equals to $594.899s$ for $K=200$, that is an increase of $61.86\%$ of speed up overall. But at $K = 5000 $ the overall simulation increases to $1100.208s$, an increase of $12.48\%$ compare to without using the Verlet buffer method.
    
    \item The most improvement is achieved with the \textsc{Avalanche} test case with a $70\%$ of simulation time improvement. The least gain is obtained with the \textsc{Biomass} test case with a $38\%$ of time improvement. The latter can be explained by the fact that it includes a \textsc{CFD} coupling with \textsc{OpenFoam}, adding therefore more computations. The percentage gain relative to broad-phase is similar than in the \textsc{Avalanche} test.
    
    The behaviour confirms the existence of an optimum $K$ value for the Verlet buffer method. The decrease in the simulation time observed at the beginning is a result of the two preceding bullet points. Undoubtedly, the values of $K$ after zero offer a larger gain in broad-phase time than the increase noticed in the narrow-phase. But after a value referenced as the optimum $K$ value, the increase in the narrow-phase time is more significant than the decrease in the broad-phase time resulting in an increase of the overall simulation time.
    
    \item At a fixed state condition, the optimum $K$ increases with the system size due to the density and cell size change. For the hopper discharge case, $K_{optimum}=100$ for $N=125k$ particles and $K_{optimum}=200$ for $N=500k$ particles.
    
    \item The simulation time does not increase at any point but stabilises rather at a minimum after a clear decrease as the interaction radius reaches the minimum cell size for the powder laser melting case.
    
    \item We can notice that our current approach of a dynamic skin gives a better performance when reaching the optimum skin. The difference is observable even for a case(biomass furnace) with a relative homogeneous flow regime.
    
    \item The optimum skin distance and $K$ value depend on the test case and therefore on several parameters as the solid fraction, the cell size, the ratio of particle to cell size and the number of particles. A study on how to compute the optimum skin distance depending on the aforementioned parameters is presented in our paper~\cite{wahidpdco}

\end{itemize}

\begin{figure}
\centering
\includegraphics[width=0.75\textwidth]{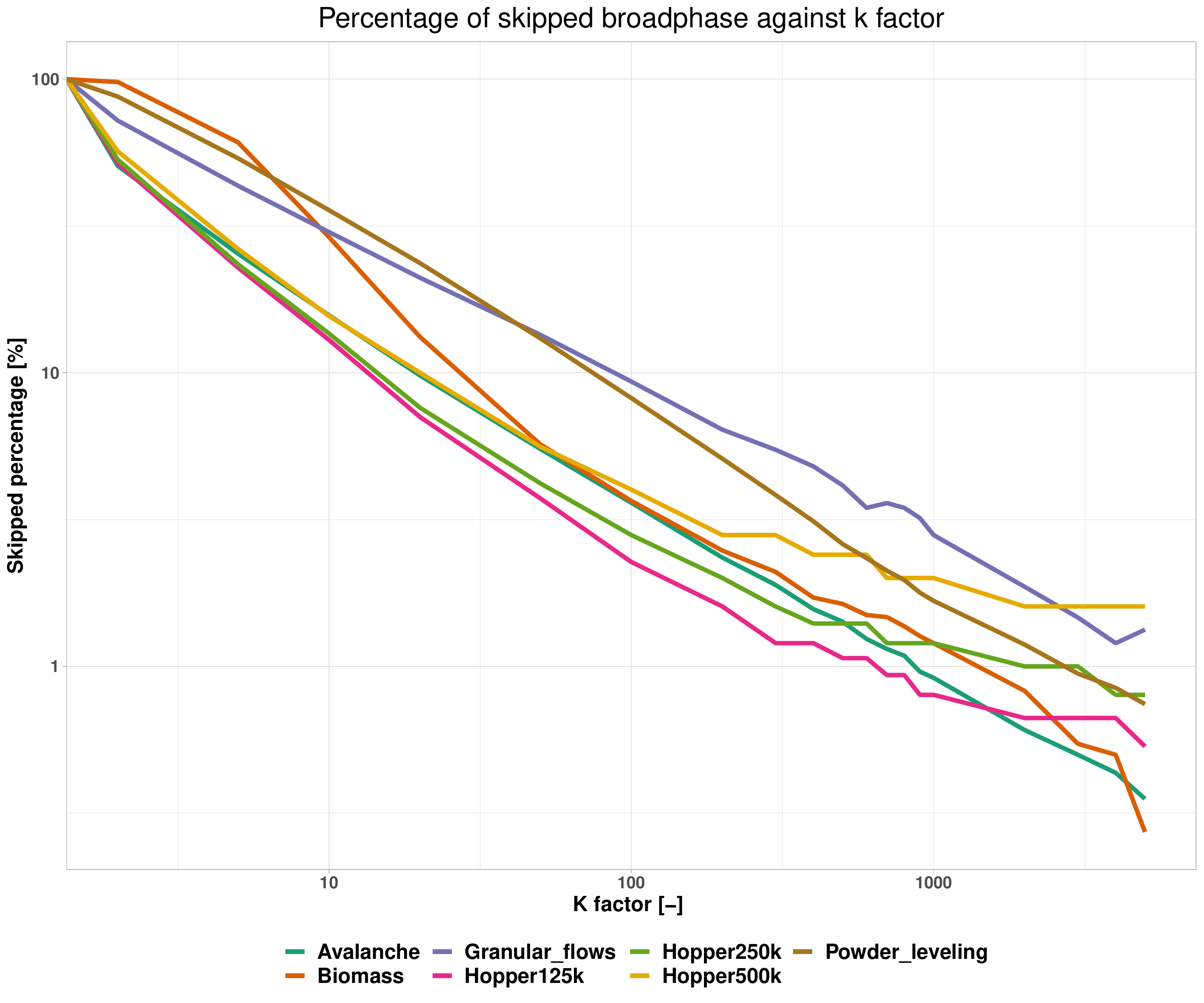}
\caption{Dependence of executed broad-phase in percentage upon the skin K factor. The percentage correspond to the number of executed broad-phase over the total number of steps in the simulation.}
\label{executed}
\end{figure}

In summary, the bigger $K$ is, the wider the interaction range is, and the wider the neighbourhood is, involving a reduction in broad-phase and simulation overall time, although the narrow-phase is increasing. Then, after $K$ reaches an optimum value, the decrease time in the broad-phase does no longer compensates the increase time in the narrow-phase leading to an increase of the overall simulation. This behaviour is observed when interaction never reaches the cell size. When it reaches the cell size, the simulation time remains unchanged since the skin margin is down to the value of the cell size.
The number of executed broad-phase as a function of $K$ value is shown in Fig.~\ref{executed}. There is a clear decrease in the number of performed broad-phase when increasing the $K$ (increase in the number of skipped broad-phase), but an equilibrium is reach around $K=200$. It means that after this value, there should be no more significant gain in skipping the broad-phase. It appears from all our simulations that the biggest drop in the simulation time is made around $K=10$ and $K=50$, although there is a clear gain by increasing the skin after those values.  


\begin{figure}[htbp]
  \centering
  \captionsetup{justification=centering}
  \includegraphics[width=0.8\linewidth]{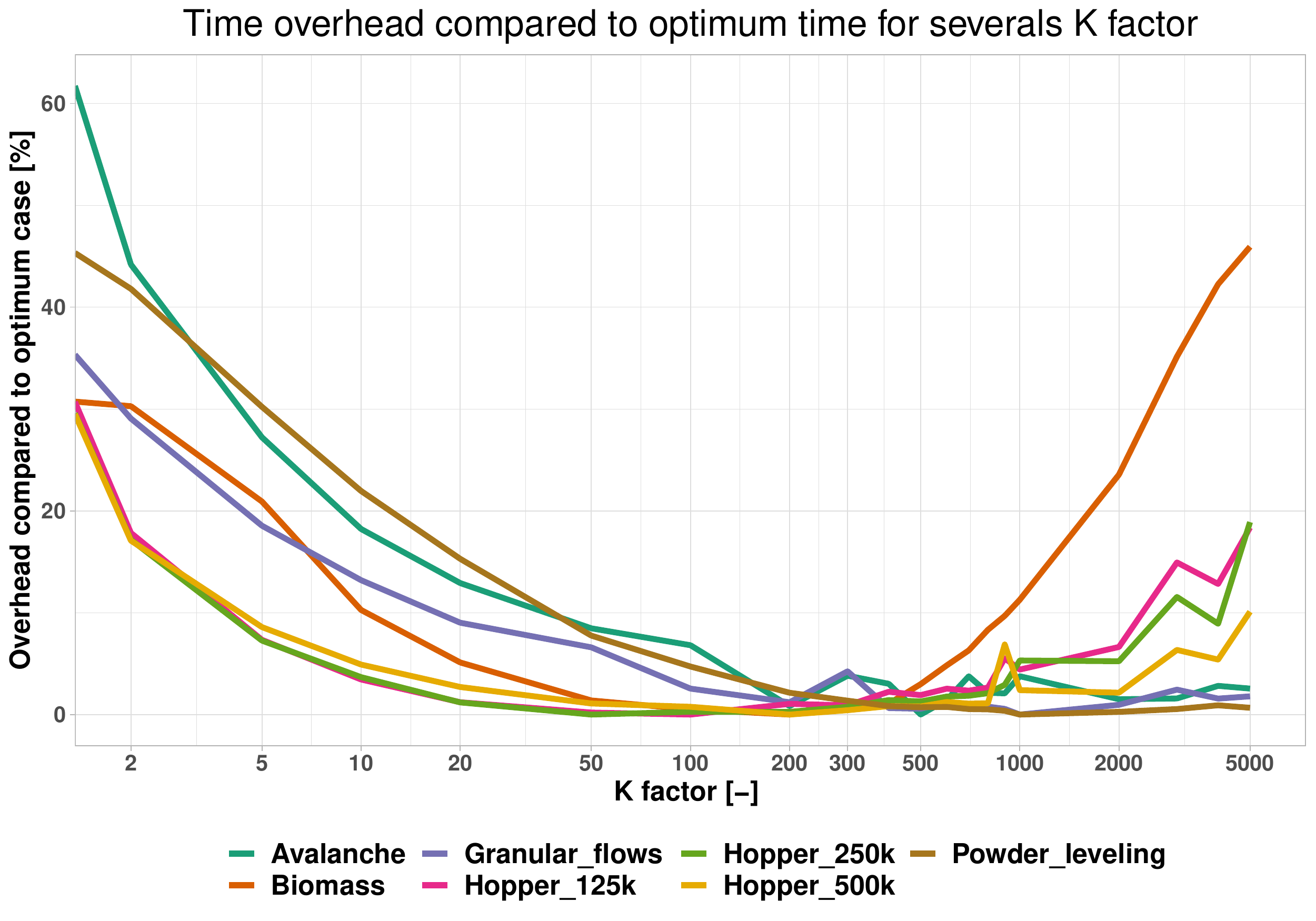}
  \caption{Simulation time overhead compared to the optimum for each $K$ value for all test cases.}
  \label{fig:overhead}
\end{figure}

\noindent Fig.~\ref{fig:overhead} enables us to put our previous observations(not significant gain after $K=200$) into perspective. It shows the time overhead for several $K$ values compared to the optimum time value. It confirms that the biggest time drop is made around $K=[10-50]$ for all the test cases. It also helps to notice that $K=200$ is a value close to the optimum for all the test cases and can thus be chosen as a common and default value.

The table~\ref{tab:summary} presents a performance comparison between the optimum case, considered as the $K$ value given the lower simulation time, which is specific to each test case, and a default case. The latter is defined as the $K$ value given an excellent performance compromise for any case. In table~\ref{tab:summary}, the improvement in percentage is defined by the gain made compared to a simulation without using the Verlet buffer method. It is given by the following formula: 

\begin{equation}
   Improvement = \frac{Time_{w/o \ Verlet \ buffer} - Time_{case} }{Time_{w/o \ Verlet \ buffer}} \times 100
   \label{eq:impr}
\end{equation}
\noindent where, $Time_{case}$ is the simulation time depending on which value of $K$ is used for the Verlet buffer method.

The \textsc{Overhead} column correspond to the time difference (percentage) between the case where the optimum $K$ value is used and the default case with the acceptable $K$ value. We can notice from table~\ref{tab:summary}, that a default $K=200$ is an excellent compromise to the optimum $K$ value that can be used for all the test cases. Actually, it has a maximum overhead of $2\%$ when used for all the test cases. We then recommend to choose $K=200$ when using the Verlet buffer method as a good arrangement to the optimum value.

\begin{sidewaystable}
\caption{Summary of the performance results of the Verlet buffer method over the different testcases.}
\centering
\small
\def\arraystretch{1.8}
\begin{tabular}{|l||>{\raggedleft\arraybackslash}p{2.5cm}||>{\raggedleft\arraybackslash}p{1.5cm}|>{\raggedleft\arraybackslash}p{2cm}|>{\raggedleft\arraybackslash}p{2.5cm}||>{\raggedleft\arraybackslash}p{2cm}|>{\raggedleft\arraybackslash}p{2.5cm}|>{\raggedleft\arraybackslash}p{2.5cm}|}
\hline 
 & \multirow{1}{=}{\centering \bf Without Verlet buffer} 
 & \multicolumn{3}{c||}{\centering \bf Optimum Value for K} 
 & \multicolumn{3}{c|}{\centering \bf Selected Default Value $K=200$}  \\

Testcase & 
\centering Simulation time $[s]$ & 
\centering $K [-]$ & \centering Simulation time $[s]$ & \centering Improvement $[\%]$ & 
\centering Simulation time $[s]$  & \centering Improvement $[\%]$ & Overhead to optimum $[\%]$ \\
  \hline
  Avalanche & $5595.77$ & $500$ & $1673.18$ & $70.12$  & $1687.24$ & $\pmb{69.84}$ & $0.83$ \\
  \hline  
  
  Biomass & $962.90$ & $200$ & $594.89$ & $38.21$ & $594.89$ & $\pmb{38.21}$ & $0.00$ \\
  \hline  
  
  Granular flows & $2084.74$ & $1000$ & $1191.64$ & $42.83$ & $1206.19$ & $\pmb{42.14}$ & $1.20$ \\
  \hline  
  
  Hopper $125k$ & $1164.40$ & $100$ & $694.14$ & $40.38$ & $701.52$ & $\pmb{39.75}$ & $1.05$ \\
  \hline  
  
  Hopper $250k$ & $1564.46$ & $50$ & $943.14$ & $39.71$ & $945.67$ & $\pmb{39.55}$ & $0.26$ \\
  \hline  
  
  Hopper $500k$ & $1614.37$ & $200$ & $975.27$ & $39.58$ & $975.27$ & $\pmb{39.58}$ & $0.00$ \\
  \hline  
  
  Powder leveling & $733.98$ & $1000$ & $367.22$ & $49.96$ & $375.25$ & $\pmb{48.87}$ & $2.14$ \\
  \hline  
\end{tabular}
\label{tab:summary}
\end{sidewaystable}

%% file: conclusion.tex
In this article, we proposed a local Verlet buffer solution with a new skin formulation which expresses the skin margin for each particle according to the conditions of the neighbourhood flow and based on the particle velocity.
The method have also been implemented in our home software with an automatic update scheme for DEM simulations of granular material. It is an improvement and a generalisation of the conventional Verlet list method for DEM simulations. The method allows to keep for several time steps the potentially interacting pairs of particles list, by surrounding the cut-off radius of particle by a skin margin, in case  where the contact detection is divided into two steps: the broad and narrow phases. It has the advantage of giving the possibility to use any contact detection algorithm to generate the approximate interacting pairs of particles list during the broad-phase process.

We presented a new skin margin formulation based on individual-particle displacements for an easy implementation and to better take into account the different flow velocity regimes that often coexist in granular flow DEM simulations. A parameter study on the skin margin was conducted in order to asses its effects on the method performances. It appears as expected, a decrease in the broad-phase overall time and an increase in the narrow-phase time while increasing the skin margin, resulting in a global decrease of the simulation time. Beyond certain values of the skin margin, we found an opposite effect where the increase in the narrow-phase time is to high resulting in a global increase of the simulation time. The study showed that most of computational time gain is made around $K=20$, and there is often after that value some gains to make, but not as much significant.